\documentclass{article}

\usepackage{authblk}

\usepackage[utf8]{inputenc}

\usepackage{graphicx}
\usepackage{amsmath, amssymb}

\usepackage{multirow} 
\newcommand{\mat}[1]{\mathbf{#1}}

\newcommand\ii{\mathrm{i}}

\newcommand\nm{\mathrm{~nm}}
\newcommand\sinv{\mathrm{~s}^{-1}}

\title{How many parameters do we need to obtain \\a BIC in symmetric and asymmetric structures?}

\author{Dmitry A. Bykov}
\author{Evgeni A. Bezus}
\author{Albert A. Mingazov}
\author{Leonid L. Doskolovich}

\affil{{\small Image Processing Systems Institute --- Branch of the Federal Scientific Research Centre ``Crystallography and Photonics'' of Russian Academy of Sciences, \authorcr 151 Molodogvardeyskaya st., Samara 443001, Russia\authorcr\;\vspace{-3ex}}}
\affil{\small Samara National Research University, \authorcr 34 Moskovskoye shosse, Samara 443086, Russia}

\date{\vspace{-5ex}}

\begin{document}
\maketitle
\begin{abstract}
Photonic bound states in the continuum (BICs) are nonradiating eigenmodes of structures with open scattering channels.
Most often, BICs are studied in highly symmetric structures with one open scattering channel.
In this simplest case, the so-called symmetry-protected BICs can be found by tuning a single parameter, which is the light frequency.
Another kind of BICs---accidental BICs---can be obtained by tuning two parameters.
For more complex structures lacking certain symmetries or having several open scattering channels, more than two parameters might be required.
In the present work, we propose an algebraic approach for computing the number of parameters required to obtain a BIC by expressing it through the dimension of the solution set of certain algebraic equations.
Computing this dimension allows us to relate the required number of parameters with the number of open scattering channels without solving Maxwell's equations.
We show that different relations take place when the scattering matrices describing the system are symmetric or asymmetric.
The obtained theoretical results are confirmed by the results of rigorous electromagnetic simulations.
\end{abstract}

\section{Introduction}
In photonics, there is always a tradeoff between the symmetry of the structure and the number of parameters required to obtain desired optical properties. 
The more symmetric the structure, the fewer parameters we need to tune.
For example, a reflection zero in a lossless layered structure can be obtained by tuning a single parameter, e.\,g., the wavelength, provided that the structure possesses a horizontal symmetry plane.
This is the case for Fabry--P\'erot resonators or phase-shifted Bragg gratings in a symmetric environment.
If, however, we consider a structure \emph{without} a horizontal symmetry plane, achieving zero reflection is still possible yet requires tuning \emph{two} parameters.
For example, in the case of a conventional antireflection coating, both the refractive index and the thickness have to be tuned in order to eliminate reflection at a fixed angle of incidence and a fixed wavelength.
In this paper, we study the number of parameters required to obtain another important phenomenon: a bound state in the continuum (BIC).

A BIC is an eigenmode of a structure having open scattering channels, which, however, does not leak out into these channels~\cite{Hsu:2016:nrm, Koshelev:2019:Nanoph, Sadreev:2021:rpp}.
In non-absorbing structures, leakage suppression makes the lifetime and the Q-factor of such eigenmodes infinite.
In realistic cases, the Q-factor inevitably becomes finite due to imperfections in the geometry of the structure, its finite size, and the presence of non-zero absorption losses.
However, the quality factor may remain very high, leading to the appearance of the so-called quasi-BICs, which are promising for practical applications including filtering, sensing, and lasing~\cite{Azzam:2021:aom}.

The most commonly encountered kind of BICs are the symmetry-protected ones, which are
the antisymmetric eigenmodes supported by symmetric structures.
Finding such a BIC requires to tune only one parameter, namely, the light frequency.
Such BICs have been extensively studied in the existing works (see, e.\,g., Refs.~\cite{Cui:sr:2016, Bulgakov:josab:2018, Bykov:2019:pra, Shipman:prb:2005}).
Another kind of BICs, often referred to as ``accidental'' BICs, requires tuning two parameters to arise.
Although such BICs are not considered as symmetry-protected, they usually appear in highly-symmetric structures with only one independent scattering channel.
Accidental BICs have been studied in various optical structures including 
photonic crystal slabs and resonant gratings~\cite{Blanchard:2016:prb, Bulgakov:josab:2018, Bykov:2019:pra, Ovcharenko:2020:prb, Bezus:2021:n, Abdrabou:2023:pra}, 
infinite chains of spheres~\cite{Bulgakov:2017:prl},
disks~\cite{Sidorenko:2021:prappl},
and cylinders~\cite{Blanchard:2016:prb, Yuan:2017:pra, Yuan:2020:pra, Yuan:2021:pra}, and ridge resonators~\cite{Zou:2015:lpr, Bezus:2018:pr, Nguyen:2019:lpr, Bykov:2020:n}, to name a few.

If we try to obtain a BIC in a structure with several open scattering channels or in a structure lacking certain symmetries (e.\,g., the Lorentz reciprocity), we will fail to do it, as long as we limit ourselves with only two parameters.
More parameters are required for less symmetric structures.
In particular, BICs obtained by tuning three parameters have been demonstrated in~\cite{Bulgakov:josab:2018}, although it was not explicitly stated in the paper.
Given the symmetry of a structure and the number of open scattering channels, finding the number of parameters required to obtain a BIC is not an easy problem.
In the recent works~\cite{Abdrabou:2023:pra, Yuan:2020:pra, Yuan:2021:pra}, the authors developed an approach for relating the number of parameters required to obtain a BIC in a periodic structure with the number of open scattering channels.
However, the use of the equation proposed in~\cite{Abdrabou:2023:pra} is not straightforward, since one has to find a coefficient (the so-called number of real constraints) corresponding to each scattering channel, which has to be obtained by careful analysis of each particular structure. 

In the present work, we propose an algebraic approach for finding the number of parameters required to obtain a BIC.
This approach can be applied to a wide range of photonic structures, which can be represented as two or more coupled scatterers with a finite number of scattering channels, and, therefore, can be described using the scattering matrix formalism.
We show that the number of parameters is related with the dimensions of two algebraic varieties being the solution sets of systems of algebraic equations.
These equations capture the symmetry properties of the structure and can be easily obtained without solving Maxwell's equations.
We show that the dimensions of the algebraic varieties can be efficiently computed using a simple algorithm.
In contrast to the approach of~\cite{Abdrabou:2023:pra, Yuan:2020:pra, Yuan:2021:pra}, the proposed method does not require performing a ``manual'' analysis of each structure and, once implemented, can be applied ``automatically'' provided that the form of the scattering matrices describing the structure is known.
By studying structures possessing different particular symmetries, we establish simple closed-form expressions relating the number of parameters with the number of open scattering channels.
Moreover, we show that once asymmetric scattering matrices are involved in describing the structure, the number of parameters may become dependent not only on the number of open scattering channels, but also on the number of modes propagating inside the structure and coupling the scatterers.
We show that the results of the proposed algebraic approach for estimating the number of parameters agrees with the results of rigorous simulations based on the numerical solution of Maxwell's equations.
We believe that the proposed approach is important for the design of structures supporting BICs as well as for understanding, which particular symmetries should be exploited to obtain a BIC with a reasonable number of parameters.

\section{Bound states in the continuum and dimensions of certain algebraic varieties}

\subsection{Multiple interference model}
Multiple interference model is an efficient approach for describing BICs in various photonic structures~\cite{Bezus:2018:pr, Bykov:2019:pra, Bykov:2020:n, Ovcharenko:2020:prb, Bezus:2021:n}.
In this paper, we will rely on a generalized version of this model to 
answer the question posed in the title.

\begin{figure}[t]
	\centering
		\includegraphics{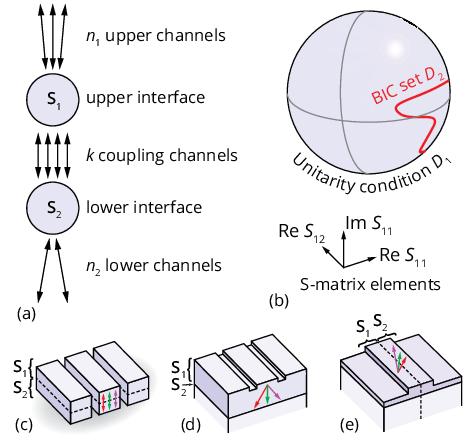}
	\label{fig:1}
	\caption{(a)~Model of a structure supporting a BIC: two coupled scatterers with S-matrices $\mat{S}_1$ (upper interface) and $\mat{S}_2$ (lower interface).
	(b)~The S-matrix parameter space (shown in 3D) containing a set of unitary matrices (shown as a 2D sphere) and a set corresponding to the structures supporting BICs (shown with a 1D red line).
	Sample structures that can be described with the proposed model: high-contrast grating~(c), guided-mode resonant grating~(d), and integrated ridge resonator~(e).}
\end{figure}

To formulate the model, let us assume that the considered structure can be represented as two coupled scatterers, as it is shown in Fig.~\ref{fig:1}(a).
The upper scatterer has $n_1$ open scattering channels and is described by the scattering matrix $\mat{S}_1$.
The lower scatterer with the S-matrix $\mat{S}_2$ has $n_2$ open scattering channels.
Besides, we assume that there are $k$ channels connecting the scatterers, which we will further refer to as the coupling channels.
Therefore, the upper scattering matrix $\mat{S}_1$ has the size $(n_1+k) \times (n_1+k)$, and the size of the lower matrix $\mat{S}_2$ is $(n_2+k) \times (n_2+k)$.

The presented model can be applied for describing various photonic structures; several particular structures that can be represented as a pair of coupled scatterers are shown in Figs.~\ref{fig:1}(c)--(e).
In what follows, we will refer to these scatterers as the interfaces of the structure.
Such interfaces may be conventional planes separating different materials [see the bottom interface in Fig.~\ref{fig:1}(d)], interfaces between homogeneous and stratified media [see Fig.~\ref{fig:1}(c)], diffraction gratings [see Fig.~\ref{fig:1}(d)], or parts of ridges and grooves of integrated optical structures [see Fig.~\ref{fig:1}(e)].
It is worth noting that usually, the scatterers constituting the structure are assumed to be located at a certain distance from each other, so that an additional diagonal scattering matrix has to be introduced to describe the phase shifts acquired by the waves (corresponding to the coupling channels) upon propagation between the scatterers (interfaces).
However, in the present work, we assume these phase shifts to be taken into account in the elements of the matrices $\mat{S}_1$ and/or $\mat{S}_2$.

Let us assume that the scattering channels of the upper interface numbered $1\ldots k$ are the coupling channels and the remaining ones are the open ``upper'' channels [see Fig.~\ref{fig:1}(a)].
In this case, we can write the upper S-matrix in the following block matrix form:
$$
\mat{S}_1=
\begin{bmatrix}
\mat{R}_1 & \mat{t}_1
\\
\mat{t}'_1 & \mat{r}_1
\end{bmatrix},
$$
where $\mat{R}_1$ is the $k\times k$ sub-matrix describing how the complex amplitudes of the waves propagating in the coupling channels change upon reflection from the upper interface.
A similar notation will be used for the lower interface:
$$
\mat{S}_2 =
\begin{bmatrix}
\mat{R}_2 & \mat{t}_2
\\
\mat{t}'_2 & \mat{r}_2
\end{bmatrix}.
$$

Since BICs appear only in lossless structures, in what follows, we will assume that the matrices $\mat{S}_1$ and $\mat{S}_2$ are always unitary (i.\,e., $\mat{S}_1^\dagger\mat{S}_1=\mat{I}$ and $\mat{S}_2^\dagger\mat{S}_2=\mat{I}$, where $\mat{I}$ is the identity matrix).

\subsection{Bound states in the continuum}

Assuming that in the studied structures [see Fig.~\ref{fig:1}(a)], BICs emerge due to reflections between the interfaces (scatterers), we can easily write down the condition of the BIC formation.
To do this, we write the round-trip operator $\mat{R}_1 \mat{R}_2$, which describes how the amplitudes of waves in the coupling channels change after two reflections (from the lower and then the upper interfaces)~\cite{Karagodsky:2011:ol}.
If there exists an amplitude vector $\mat{d}\in\mathbb{C}^k$, which does not change after these reflections (i.\,e., $\mat{R}_1 \mat{R}_2\mat{d} = \mat{d}$), then the considered structure supports a BIC.
Obviously, this is possible if and only if the round-trip operator has a unit eigenvalue.
This allows us to write the following BIC condition~\cite{Bezus:2018:pr}:
\begin{equation}
\label{eq:BIC}
\det \left(\mat{R}_1\mat{R}_2-\mat{I}\right) = 0.
\end{equation}

\subsection{Dimensions and parameters}
In the proposed approach, we will consider the elements of the matrices $\mat{S}_1$ and $\mat{S}_2$ as variables.
For symmetric structures, one such variable can occur multiple times in one or both matrices.
Assuming that the total number of \emph{unique} variables equals $m$, we deduce that a structure described by such pair of matrices corresponds to a point in the space $\mathbb{R}^{2m}$, where the factor 2 is due the fact that one has to define real and imaginary parts of each variable.

Let us now consider a set of all \emph{unitary} matrices $\mat{S}_1$ and $\mat{S}_2$:
\begin{equation}
\label{eq:D1}
D_1 = \left\{
\begin{aligned}
&\mat{S}_1^\dagger\mat{S}_1=\mat{I},
\\
&\mat{S}_2^\dagger\mat{S}_2=\mat{I}.
\end{aligned}
\right.
\end{equation}
Here, we use the notation ``$D_1 = \{\cdots$'' meaning that $D_1$ is the set of solutions of a system of algebraic equations.
The explicit form of these equations, which are written in matrix form in Eq.~\eqref{eq:D1}, is presented in the Supplementary.
Note that $D_1$ is a subset of $\mathbb{R}^{2m}$.

Next, let us define the set $D_2$ that is a subset of $D_1$ where BICs exist.
To do this, we add the BIC condition~\eqref{eq:BIC} to the system of equations in Eq.~\eqref{eq:D1}:
\begin{equation}
\label{eq:D2}
D_2 = \left\{
\begin{aligned}
&\mat{S}_1^\dagger\mat{S}_1=\mat{I},
\\
&\mat{S}_2^\dagger\mat{S}_2=\mat{I},
\\
&\det \left(\mat{R}_1\mat{R}_2-\mat{I}\right) = 0.
\end{aligned}
\right.
\end{equation}
Since the sets $D_1$ and $D_2$ are both defined by systems of algebraic equations, they can be referred to as algebraic varieties.
An example of a system defining the variety $D_2$ is also given in the Supplementary.

Figure~\ref{fig:1}(b) illustrates the idea behind finding the number of parameters required to obtain a BIC.
In this figure, the coordinate axes (only three are shown) correspond to the parameters defining the S-matrices.
Let us remind that these parameters are the real and imaginary parts of the unique elements of the S-matrices.
The condition for the S-matrices to be unitary defines a certain region $D_1$ in the considered space.
For illustrative purposes, it is shown as a 2D sphere in the figure.
In the general case, $D_1$ is an algebraic variety having a certain dimension $\dim D_1$.
A set $D_2$ of unitary scattering matrices satisfying the BIC condition~\eqref{eq:BIC} is a subset of $D_1$ and is shown with a 1D red line on the sphere.

If we want to obtain a BIC, we have to tune several parameters (e.\,g.,~$N$ parameters), which are assumed independent and having no hidden symmetries.
Besides, we assume that these parameters are real and changing them preserves the unitarity of the S-matrices (i.\,e., keeps us inside the~$D_1$ set).
Therefore, varying the values of these parameters allows us to cover a certain $N$-dimensional region on the ``sphere''~$D_1$.
In this case, the ``probability'' of finding a BIC can be non-zero only if the dimension of the BIC set, which is $\dim D_2$, and the dimension of the ``search area'', which is $N$, add up to at least the dimension of the ``unitarity set''~$D_1$.
Therefore, we can calculate the minimum number of parameters required to obtain a BIC as the difference between the dimensions of the varieties~$D_1$ and~$D_2$:
\begin{equation}
\label{eq:D}
N = \dim D_1 - \dim D_2.
\end{equation}
For the illustration presented in Fig.~\ref{fig:1}(b), we require $N=1$ parameter; tuning this parameter would allow us to cover a line on the sphere intersecting with the red BIC line with a non-zero probability.


In what follows, we will use Eq.~\eqref{eq:D} to calculate the number of parameters required to obtain a BIC.
To do this, we have to calculate the dimensions of the solution sets (or, strictly speaking, real dimensions of algebraic varieties) $D_1$ and $D_2$,
which can be done using the so-called cylindrical algebraic decomposition (CAD) or other numerical methods~\cite{Basu:2006}.
We describe the particular method we used to compute these dimensions and some implementation details in the Supplementary.

\section{Computing the number of parameters required to obtain a BIC}

In this section, we consider several particular forms of the scattering matrices $\mat{S}_1$ and $\mat{S}_2$ and calculate the corresponding number of parameters~$N$ using Eq.~\eqref{eq:D}.

\subsection{BICs in symmetric reciprocal structures}\label{ssec:theory:sym}
Let us first consider the case when both $\mat{S}_1$ and $\mat{S}_2$ are symmetric.
This holds for structures satisfying the Lorentz reciprocity conditions, in particular, for structures made of ``conventional'' non-magnetized dielectric materials.
Thus, this case covers the majority of the existing works dedicated to the invesigation of BICs.
It is important to note that if gratings are considered as interfaces (scatterers), a vertical symmetry plane is also required for the scattering matrix to be symmetric~\cite{Gippius:prb:2005}.

For symmetric matrices $\mat{S}_1$ and $\mat{S}_2$, the dimension of the variety $D_1$ is easily calculated as~\cite{Dita:jpa:1982}
\begin{equation}
\label{eq:dimD1sym}
\dim D_1 = \frac{(n_1+k)^2+n_1+k}2 + \frac{(n_2+k)^2+n_2+k}2,
\end{equation}
where the first and the second terms describe the quantity of independent real numbers required to define the upper and lower unitary scattering matrices, respectively.
Let us note that for structures possessing a horizontal symmetry plane, for which $\mat{S}_1 = \mat{S}_2$, one should keep only the first term in Eq.~\eqref{eq:dimD1sym}. 
Finding the dimension of the BIC variety $D_2$ analytically is much more complicated.
In this regard, we utilized a numerical approach described in the Supplementary for computing $\dim D_2$.

Having calculated the dimensions $\dim D_1$ and $\dim D_2$, we find the number of parameters required to obtain a BIC using Eq.~\eqref{eq:D} and denote it by $N_{\rm sym}$.
The computation results are presented in Table~\ref{tab:1}.
In this table, we summarized the results for various considered structures with different numbers of upper and lower scattering channels $n_1$ and $n_2$ and coupling channels $k$.
By analyzing Table~\ref{tab:1}, one can deduce that the number of parameters required to obtain a BIC is
\begin{equation}
\label{eq:Dsym}
N_{\rm sym} = n_1+n_2+1,
\end{equation}
which depends only on the total number of independent open scattering channels ($n_1+n_2$).
Interestingly, the number of parameters does not depend on the number of coupling channels~$k$, i.\,e., on the number of waves propagating inside the structure [see Fig.~\ref{fig:1}(a)].
As shown in Table~\ref{tab:1}, we verified this for the cases of up to at least three coupling channels.
Considering larger values of $k$ than the ones presented in Table~\ref{tab:1} would require enormous amount of time when computing $\dim D_2$.

\begin{table}
	\centering
		\begin{tabular}{cccccc}
		\multicolumn{4}{c}{Considered structures}  & \multirow{2}{*}{$N_{\rm sym}$} & \multirow{2}{*}{$N_{\rm asym}$}  \\ 
		  $n_1$ & $n_2$ & $k$ & $n_1+n_2$ & &\\ \hline	
			
1 & ---  & 1\ldots6 & \multirow{2}{*}1 & \multirow{2}{*}2 & \multirow{2}{*}3 \\ 
1 & 0 & 1\ldots3 & & \\  \hline

2 & --- & 1\ldots4 & \multirow{3}{*}2 & \multirow{3}{*}3  & \multirow{3}{*}5\\
2 & 0 & 1\ldots3 & \\
1 & 1 & 1\ldots3 & \\  \hline

3 & --- & 1\ldots3 & \multirow{3}{*}3 & \multirow{3}{*}4  & \multirow{3}{*}7\\
3 & 0 & 1\ldots3 & \\
2 & 1 & 1\ldots2 & \\  \hline

4 & --- & 1\ldots4 & \multirow{4}{*}4 & \multirow{4}{*}5  & \multirow{4}{*}9\\
4 & 0 & 1\ldots2 &  \\
3 & 1 & 1\ldots2 &  \\
2 & 2 & 1\ldots3 & \\  \hline
		\end{tabular}
	\caption{Number of parameters required to obtain a BIC in symmetric ($N_{\rm sym}$) and asymmetric/nonreciprocal ($N_{\rm asym}$) structures. Em dashes in the second column denote structures with $\mat{S}_1 = \mat{S}_2$.}
	\label{tab:1}
\end{table}

\subsection{BICs in asymmetric and nonreciprocal structures}\label{ssec:theory:asym}
When both upper and lower parts of the structure contain nonreciprocal materials or when the (reciprocal) grating under consideration has an asymmetrical unit cell, 
the scattering matrices $\mat{S}_1$ and $\mat{S}_2$ are no longer symmetric and the number of real parameters required for defining such unitary matrices is~\cite{Dita:jpa:1982}
\begin{equation}
\label{eq:dimD1asym}
\dim D_1 = (n_1+k)^2 + (n_2+k)^2.
\end{equation}
Similarly to Eq.~\eqref{eq:dimD1sym}, only one term remains in Eq.~\eqref{eq:dimD1asym} when $\mat{S}_1 = \mat{S}_2$.

By finding the dimension of the variety $D_1$ using Eq.~\eqref{eq:dimD1asym} and by calculating the dimension of~$D_2$ numerically, we filled in the last column of Table~\ref{tab:1}.
According to the table, the number of parameters required to obtain a BIC in an asymmetric/nonreciprocal structure equals
\begin{equation}
\label{eq:Dasym}
N_{\rm asym} = 2(n_1+n_2)+1,
\end{equation}
where $n_1+n_2$ is the total number of independent open scattering channels. 
As in  the previous subsection, the number of parameters $N=N_{\rm asym}$ does not depend on the number of coupling channels $k$. 

\subsection{BICs arising due to coupling of symmetric and asymmetric scatterers}
The results presented in the previous two subsections suggest that the number of parameters required to obtain a BIC does not depend on the number of coupling channels $k$ linking the upper and lower parts of the structure.
Here, we show that it is generally not the case and~$N$ might actually depend on~$k$.
To do this, let us consider a structure with a symmetric matrix $\mat{S}_1$ and an asymmetric matrix $\mat{S}_2$.
We can imagine such a structure as one having nonreciprocal materials localized only in its lower part.
For such structures, the dimension of the variety~$D_1$ is calculated analytically as $\dim D_1 = [(n_1+k)^2+n_1+k]/2 + (n_2+k)^2$ and $\dim D_2$ is calculated numerically.

The calculation results are presented in Table~\ref{tab:2}, according to which the number of parameters indeed becomes dependent on the number of coupling channels $k$.
The general equation describing the numerical evidence presented in Table~\ref{tab:2} reads as
\begin{equation}
\label{eq:Dmix}
N = n_1 + \min\{n_1+1, k\} + 2 n_2,
\end{equation}
where $n_1$ is the number of the open scattering channels of the symmetric (reciprocal) part of the structure and~$n_2$ is that of the asymmetric (nonreciprocal) part.
Interestingly, at~$k=1$, the upper and lower parts are coupled by only one scattering channel, so that the reciprocal upper part ``does not feel'' the nonreciprocity in the lower part.
As a consequence, the required number of parameters becomes $N = n_1 + 2 n_2  + 1$, which includes $n_1$ with a factor of 1 as in Eq.~\eqref{eq:Dsym} and $n_2$ with a factor of 2 as in Eq.~\eqref{eq:Dasym}.
On the other hand, when the number of coupling channels~$k$ is sufficiently large ($k > n_1$), the whole structure can be treated as an asymmetric (nonreciprocal) one and Eq.~\eqref{eq:Dmix} turns into the corresponding Eq.~\eqref{eq:Dasym}.

\begin{table}
	\centering
		\begin{tabular}{ccc|ccc}
		  \multirow{2}{*}{$n_1+n_2$} & \multirow{2}{*}{$n_1$} & \multirow{2}{*}{$n_2$} & 
			\multicolumn{3}{c}{$N$}\\
			 & & & $k=1$&$k=2$&$k=3$\\\hline
\multirow{2}{*}1	
& 1 & 0 & 2 & 3 & 3 \\
& 0 & 1 & 3 & 3 & 3    \\ \hline
\multirow{3}{*}2	
& 2 & 0 & 3 & 4 & 5    \\
& 1 & 1 & 4 & 5 & 5    \\
& 0 & 2 & 5 & 5 & 5    \\ \hline
\multirow{4}{*}3	
& 3 & 0 & 4 & 5 & 6  \\
& 2 & 1 & 5 & 6 & 7  \\
& 1 & 2 & 6 & 7 & 7  \\
& 0 & 3 & 7 & 7 & 7  \\ \hline
\multirow{5}{*}4	
& 4 & 0 & 5 & 6 & 7  \\
& 3 & 1 & 6 & 7 & 8 \\
& 2 & 2 & 7 & 8 & 9  \\
& 1 & 3 & 8 & 9 & 9  \\
& 0 & 4 & 9 & 9 & 9  \\ \hline		\end{tabular}
	\caption{Number of parameters $N$ required to obtain a BIC in structures with coupled symmetric (reciprocal) and asymmetric (nonreciprocal) parts.}
	\label{tab:2}
\end{table}

\section{Numerical examples}
In this section, we consider several examples demonstrating that the predictions of the proposed algebraic approach agree with the rigorous simulation results obtained by numerically solving Maxwell's equations.

\subsection{1-parameter symmetry-protected BIC in a GMRG}
First, let us consider a symmetric reciprocal guided-mode resonant grating [i.\,e., a grating on a slab waveguide shown in Fig.~\ref{fig:1}(d)], which is known for supporting symmetry-protected BICs~\cite{Cui:sr:2016, Bulgakov:josab:2018, Bykov:2019:pra}.
Typical structures considered in Refs.~\cite{Cui:sr:2016, Bulgakov:josab:2018, Bykov:2019:pra} have only the zeroth diffraction orders propagating in the superstrate and substrate, hence we assume $n_1 = n_2 = 1$ in the developed model.
Inside the waveguide layer, there are usually three propagating diffraction orders having the numbers $-1,0,1$, which gives $k=3$ coupling channels.
To apply the proposed model, we will treat the diffraction grating with the waveguide layer as the upper scatterer (interface) of the structure having the scattering matrix~$\mat{S}_1$.
The lower scatterer~$\mat{S}_2$ in this case will describe only the bottom interface of the waveguide layer [see Fig.~\ref{fig:1}(d)].
This gives us the following form of the scattering matrices (see, e.\,g.,~\cite{Bykov:2019:pra} for details):
$$
\mat{S}_1 =
\begin{bmatrix}
r_m & b & d_2 & d_1 \\
b & r_m & d_2 & d_1 \\
d_2 & d_2 & r_3 & t_2 \\
d_1 & d_1 & t_2 & r_4
\end{bmatrix},
\;\;\;
\mat{S}_2 = 
\begin{bmatrix}
\rho & 0 & 0 & 0 \\
0 & \rho & 0 & 0 \\
0 & 0 & r_1 & t_1 \\
0 & 0 & t_1 & r_2
\end{bmatrix},
$$
the elements of which are complex numbers.
Note that the exponents describing the propagation of the plane waves corresponding to the coupling channels between the interfaces of the waveguide layer, which usually appear in multiple interference models, are assumed to be incorporated into the elements of the scattering matrix $\mat{S}_1$.

It is important to note that, due to the presence of additional symmetries, the form of the scattering matrices differs from the general form considered in the previous section.
Therefore, we cannot use Eq.~\eqref{eq:dimD1sym} or~\eqref{eq:dimD1asym} to calculate the dimension of the set of unitary matrices $D_1$; thus, this dimension has to be computed numerically, which gives $\dim D_1 = 11$.
Similarly, for the considered example, the numerically found dimension of the BIC set amounts to $\dim D_2 = 10$.
Therefore, according to Eq.~\eqref{eq:D}, $N = 1$ and the BIC in the considered structures requires tuning a single parameter to arise.
This is indeed the case for symmetry-protected BICs appearing at normal incidence in resonant guided-mode gratings and photonic crystal slabs.
As it was mentioned above, it is the frequency of the incident light, which is usually used as such a parameter.
Since symmetry-protected BICs have been vastly studied in the literature~\cite{Cui:sr:2016, Bulgakov:josab:2018, Bykov:2019:pra, Shipman:prb:2005}, we do not present simulation results for this case and move to more interesting examples.

\subsection{2-parameter BIC in a symmetric HCG}\label{ssec:sym}

\begin{figure}[t]
		\hspace{-4em}\includegraphics[scale=0.75]{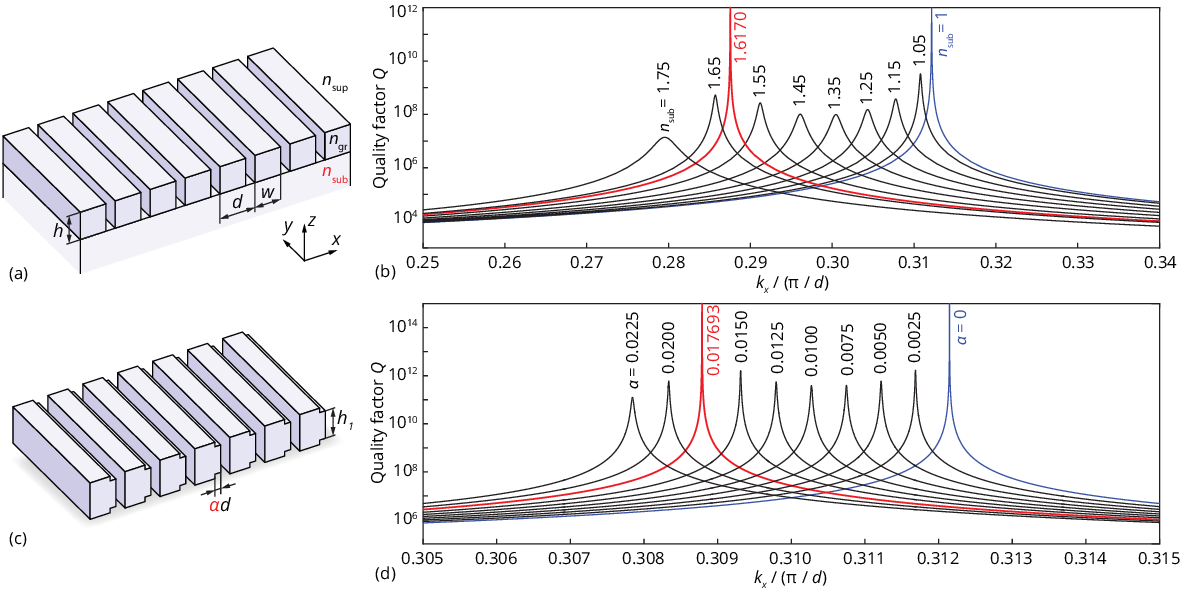}
	\label{fig:2}
	\caption{High-contrast grating on a substrate breaking the horizontal symmetry~(a)
and the quality factor of its eigenmodes vs.\ the in-plane wave vector component $k_x$ at different values of $n_{\rm sub}$~(b).
High-contrast grating with broken vertical symmetry~(c)
and the quality factor of its eigenmodes vs.\ $k_x$ at different values of $\alpha$~(d).
Blue lines in (b) and (d) correspond to the symmetric structure supporting a BIC; red lines demonstrate the appearance of 3-parameter BICs when the symmetry is broken.}
\end{figure}

As the second example, we consider a suspended high-contrast grating shown in Fig.~\ref{fig:1}(c) with the period  $d = 800\nm$, thickness $h = 900\nm$, fill factor $w/d = 0.8$, and the following values of the refractive indices of the grating and the surrounding medium: $n_{\rm gr} = 3.5$ and $n_{\rm sup} = n_{\rm sub} = 1$ [see Fig.~\ref{fig:2}(a)].
We will focus on the case of planar diffraction ($k_y = 0$) and the transverse electric (TE) polarization of light.

This grating supports a BIC, which can be found by tuning the angular frequency $\omega$ and the in-plane wave number $k_x$~\cite{Ovcharenko:2020:prb}.
This is demonstrated by the blue line in Fig.~\ref{fig:2}(b), which shows the dependence of the quality factor of the eigenmode along its dispersion curve as we change the value of the in-plane wavenumber~$k_x$.
To calculate the Q-factor, we used the numerical method of paper~\cite{Bykov:2013:jlt}, which is based on the Fourier modal method (FMM), an established numerical technique for solving Maxwell's equations in the case of periodic structures~\cite{Moharam:1995:josaa, Li:1996:josaa}.
It is evident that at a certain point, the Q-factor of the mode diverges.
Therefore, we have to find a proper value of $k_x$ and then, having this value fixed, to choose a frequency $\omega$, at which a BIC appears.
This gives the total of two parameters required for obtaining a BIC.
The numerical values of these parameters calculated for the considered example using the FMM are
$$
(k_x, \omega) = (0.31218\,\pi/d,\; 1.04411\cdot10^{15}\sinv).
$$

Now, let us verify the required number of parameters using the proposed algebraic approach.
The considered HCG can be treated as an $x$-periodic one-dimensional photonic crystal truncated in the $z$ direction.
At the parameters given above, such a periodic medium supports three propagating TE-polarized modes.
Therefore, we use $k = 3$ in the proposed model.
For the considered $k_x$ range [see Fig.~\ref{fig:2}(b)], the structure has only one open scattering channel above and below the structure corresponding to the zeroth diffraction orders, therefore, we take $n_1 = 1$.
Since the structure possesses a horizontal symmetry plane, the lower scattering matrix coincides with the upper one: $\mat{S}_1 = \mat{S}_2$.
Note that this scattering matrix is symmetric due to Lorentz reciprocity and the presence of a vertical symmetry plane~\cite{Gippius:prb:2005}.
Calculating the dimensions of the sets $D_1$ and $D_2$ yields $\dim D_1 = [(n_1+k)^2+(n_1+k)]/2 = 10$ and $\dim D_2 = 8$.
Therefore, we indeed require $N = 10-8=2$ parameters for obtaining a BIC, which agrees with the FMM simulation results. 
We note that this case is covered by Subsection~\ref{ssec:theory:sym} and is included in Table~\ref{tab:1}.

In the next two subsections, we consider the cases, in which two different symmetries of the HCG are broken, which, as we will demonstrate, leads to an increase in the number of parameters required to obtain a BIC.

\subsection{3-parameter BIC in an HCG on a substrate}

Let us break the horizontal symmetry plane of the HCG by introducing a substrate with refractive index $n_{\rm sub} > 1$ as it is shown in Fig.~\ref{fig:2}(a).
Breaking this symmetry makes the scattering matrices $\mat{S}_1$ and $\mat{S}_2$ of the upper and lower interfaces different, which doubles the number of parameters required to define them.
Therefore, the dimension of the variety $D_1$ doubles as well and, in accordance with Eq.~\eqref{eq:dimD1sym}, amounts to $\dim D_1 = 20$.
Computing the dimension of $D_2$ gives $\dim D_2 = 17$.
Therefore, in accordance with Subsection~\ref{ssec:theory:sym}, it is $N = 20-17=3$ parameters, which are required to obtain a BIC in an HCG on a substrate.

This agrees with the simulation results presented in Fig.~\ref{fig:2}(b).
This figure shows that by increasing the value of $n_{\rm sub}$ from unity, one immediately destroys the BIC.
However, a further increase enables finding a ``magic'' value of $n_{\rm sub}$, at which the Q-factor diverges again [see the red line in Fig.~\ref{fig:2}(b)].
Therefore, the figure shows that \emph{three} parameters ($n_{\rm sub}$, $k_x$, and $\omega$) are required to obtain a BIC in an HCG with broken horizontal symmetry.
The FMM-calculated values of these parameters amount to
$$
(n_{\rm sub}, k_x, \omega) = (1.6170,\; 0.28762\,\pi/d,\; 1.03427\cdot10^{15}\sinv).
$$

Let us note that for illustrative purposes, in this and the following subsections, the considered examples were chosen so that tuning of the same parameter, which is used to violate the HCG symmetry, enables obtaining a BIC again.
In general, this is not necessarily the case, and another parameter (or parameters) can be used to ``restore'' a BIC in a structure with a broken symmetry.

\subsection{3-parameter BIC in an HCG with an asymmetric unit cell}

Next, let us return to the suspended structure ($n_{\rm sub} = 1$) and break the vertical symmetry of its unit cell.
To do this, we keep the total thickness of the HCG the same but change the fill factor of the upper and lower 10-nm-thick layers.
Such structure is shown in Fig.~\ref{fig:2}(c), where $h_1 = 880\nm$ and the ridge width in the 10-nm-thick layers equals $(0.8-\alpha) d$ with $\alpha$ being the symmetry breaking parameter.
At $\alpha = 0$, we obtain a symmetric HCG considered above in Subsection~\ref{ssec:sym}; increasing the $\alpha$ value leads to an increase in the asymmetry degree of the HCG.

Let us apply the proposed algebraic approach to analyze this structure.
Due to horizontal symmetry, we have $\mat{S}_1 = \mat{S}_2$, but the matrix $\mat{S}_1$ is no longer symmetric.
This is due to the fact that although the reciprocity of the structure is not violated, for gratings (or photonic crystal slabs), the presence of a vertical symmetry plane is also required for the S-matrix to be symmetric~\cite{Gippius:prb:2005}.
In accordance with Subsection~\ref{ssec:theory:asym}, for asymmetric matrices, the dimension of the unitary matrix set equals $\dim D_1 = 16$ [see Eq.~\eqref{eq:dimD1asym}].
The dimension of the BIC variety was estimated numerically and was found to be $\dim D_2 = 13$.
Therefore, three parameters are again required to obtain a BIC in this case.

This agrees with the simulation results presented in Fig~\ref{fig:2}(d).
Let us note that the ``symmetry-breaking layers'' are extremely thin (10\ nm); this makes the Q-factor of the eigenmodes to remain quite high, nevertheless, we carefully checked that the Q-factor is finite for black lines and diverges only on the plots corresponding to $\alpha = 0$ (blue line), which is the symmetric case, and to $\alpha = 0.017693$ (red line).
It is the latter case, which corresponds to a 3-parameter BIC, occurring at 
$$
(\alpha, k_x, \omega) = (0.017693,\; 0.30879\,\pi/d,\; 1.04356\cdot10^{15}\sinv).
$$

\section{Conclusion}

We proposed an algebraic approach allowing one to estimate the number of parameters required to obtain a BIC in a photonic structure without solving Maxwells's equations.
The presented approach captures the symmetry properties of the structure and reduces the calculation of the number of parameters to finding the real dimensions of two algebraic varieties.
The obtained theoretical predictions are confirmed by the rigorous numerical simulation results, with guided-mode resonant gratings and high-contrast gratings considered as examples.

We demonstrated that in the simplest case, the number of the required parameters is determined by the number of open scattering channels and by the form of the scattering matrix of the structure.
However, when one of the coupled parts of the structure is symmetric and the other is asymmetric, the behavior is more sophisticated and the number of parameters becomes dependent on the number of waves propagating inside the structure.

In the present work, we considered a model containing two coupled scatterers, however, more scattering matrices can be introduced to the model to describe optical properties of the structures possessing other symmetries.
We believe that the proposed approach can also be applied to investigate the number of parameters required to obtain other resonant phenomena, such as zeros of the reflection and transmission coefficients, total absorption, coherent perfect absorption, and lasing.

\section*{Acknowledgment}
This work was funded by the Russian Science Foundation (grant 22-12-00120).

\part*{Supplementary material}

\section{An example of algebraic equations defining the algebraic varieties $D_1$ and $D_2$}

As a matter of example, in this section we present an explicit form of the equations defining the varieties $D_1$ and $D_2$ for the case of a structure with identical upper and lower interfaces, i.\,e., with $\mat{S}_1 = \mat{S}_2$.
We will assume this scattering matrix to be symmetric.
Let us further assume that inside the structure, there are two propagating waves ($k=2$) and that there is one open scattering channel in the superstrate and substrate ($n_1 = 1$).
In this case, the scattering matrix takes the form
$$
\mat{S}_1 =
\begin{bmatrix}
s_{11} & s_{12} & s_{13} \\
s_{12} & s_{22} & s_{23} \\
s_{13} & s_{23} & s_{33}
\end{bmatrix}.
$$

Let us  write the unitarity condition $\mat{S}_1^\dagger\mat{S}_1 = \mat{I}$ for this matrix.
Since the unitarity condition contains complex conjugation, the set $D_1$ cannot be considered as a complex algebraic variety.
Therefore, we have to work with real algebraic varieties.
To do this, we rewrite each complex variable in terms of its real and imaginary parts as $s_{ij} = x_{ij} + \ii y_{ij}$. 
Then, we separately equate the real and imaginary parts in the matrix equality $\mat{S}_1^\dagger\mat{S}_1 = \mat{I}$ and obtain 
\begin{equation}
\label{eqsupp:D1}
D_1 = \left\{
\begin{aligned}
 & x_{1,1}^2 + x_{1,2}^2 + x_{1,3}^2 + y_{1,1}^2 + y_{1,2}^2 + y_{1,3}^2 = 1, \\
 & x_{1,2}^2 + x_{2,2}^2 + x_{2,3}^2 + y_{1,2}^2 + y_{2,2}^2 + y_{2,3}^2 = 1, \\
 & x_{1,3}^2 + x_{2,3}^2 + x_{3,3}^2 + y_{1,3}^2 + y_{2,3}^2 + y_{3,3}^2 = 1, \\
 & x_{1,1} x_{1,2} + x_{2,2} x_{1,2} + x_{1,3} x_{2,3} + y_{1,1} y_{1,2} + y_{1,2} y_{2,2} + y_{1,3} y_{2,3} = 0, \\
 & x_{1,2} y_{1,1} + x_{2,2} y_{1,2} + x_{2,3} y_{1,3} - x_{1,1} y_{1,2} - x_{1,2} y_{2,2} - x_{1,3} y_{2,3} = 0, \\
 & x_{1,3} y_{1,1} + x_{2,3} y_{1,2} - x_{1,1} y_{1,3} + x_{3,3} y_{1,3} - x_{1,2} y_{2,3} - x_{1,3} y_{3,3} = 0, \\
 & x_{1,3} y_{1,2} - x_{1,2} y_{1,3} + x_{2,3} y_{2,2} - x_{2,2} y_{2,3} + x_{3,3} y_{2,3} - x_{2,3} y_{3,3} = 0, \\
 & x_{1,1} x_{1,3} + x_{3,3} x_{1,3} + x_{1,2} x_{2,3} + y_{1,1} y_{1,3} + y_{1,2} y_{2,3} + y_{1,3} y_{3,3} = 0, \\
 & x_{1,2} x_{1,3} + x_{2,2} x_{2,3} + x_{2,3} x_{3,3} + y_{1,2} y_{1,3} + y_{2,2} y_{2,3} + y_{2,3} y_{3,3} = 0.
\end{aligned}
\right.
\end{equation}
Here, the notation ``$D_1 = \{\star$'' means that $D_1$ is the set of solutions of the system of equations ``$\{\star$''; strictly speaking, $D_1 = \{(x_{1,1}, \ldots, x_{3,3}, y_{1,1}, \ldots, y_{3,3}) \,|\, \star\}$.

Next, let us add the BIC condition to this system to obtain the set $D_2$.
As it was noted in the main text of the article, the BIC condition reads as $\det(\mat{R}_1\mat{R}_2-\mat{I}) = 0$.
Since in the considered case, $\mat{S}_1 = \mat{S}_2$, we can factorize the BIC condition as $\det(\mat{R}_1-\mat{I})\cdot \det(\mat{R}_1+\mat{I}) = 0$.
This form of the BIC condition allows one to study symmetric and antisymmetric BICs separately~\cite{Bezus:2018:pr}.
For the sake of simplicity, let us focus only on the symmetric BICs, which are defined by the equation $\det(\mat{R}_1-\mat{I}) = 0$.
We express the elements of the matrix $\mat{R}_1$ through their real and imaginary parts and equate separately the real and imaginary parts of the determinant to zero.
By adding these two equations to the system of Eq.~\eqref{eqsupp:D1}, we finally obtain the algebraic equations defining the BIC set~$D_2$:
\begin{equation}
\label{eqsupp:D2}
D_2 = \left\{
\begin{aligned}
 & x_{1,1}^2 + x_{1,2}^2 + x_{1,3}^2 + y_{1,1}^2 + y_{1,2}^2 + y_{1,3}^2 = 1, \\
 & x_{1,2}^2 + x_{2,2}^2 + x_{2,3}^2 + y_{1,2}^2 + y_{2,2}^2 + y_{2,3}^2 = 1, \\
 & x_{1,3}^2 + x_{2,3}^2 + x_{3,3}^2 + y_{1,3}^2 + y_{2,3}^2 + y_{3,3}^2 = 1, \\
 & x_{1,1} x_{1,2} + x_{2,2} x_{1,2} + x_{1,3} x_{2,3} + y_{1,1} y_{1,2} + y_{1,2} y_{2,2} + y_{1,3} y_{2,3} = 0, \\
 & x_{1,2} y_{1,1} + x_{2,2} y_{1,2} + x_{2,3} y_{1,3} - x_{1,1} y_{1,2} - x_{1,2} y_{2,2} - x_{1,3} y_{2,3} = 0, \\
 & x_{1,3} y_{1,1} + x_{2,3} y_{1,2} - x_{1,1} y_{1,3} + x_{3,3} y_{1,3} - x_{1,2} y_{2,3} - x_{1,3} y_{3,3} = 0, \\
 & x_{1,3} y_{1,2} - x_{1,2} y_{1,3} + x_{2,3} y_{2,2} - x_{2,2} y_{2,3} + x_{3,3} y_{2,3} - x_{2,3} y_{3,3} = 0, \\
 & x_{1,1} x_{1,3} + x_{3,3} x_{1,3} + x_{1,2} x_{2,3} + y_{1,1} y_{1,3} + y_{1,2} y_{2,3} + y_{1,3} y_{3,3} = 0, \\
 & x_{1,2} x_{1,3} + x_{2,2} x_{2,3} + x_{2,3} x_{3,3} + y_{1,2} y_{1,3} + y_{2,2} y_{2,3} + y_{2,3} y_{3,3} = 0, \\
 & x_{1,2}^2 + x_{1,1} - x_{1,1} x_{2,2} + x_{2,2} - y_{1,2}^2 + y_{1,1} y_{2,2} = 1, \\
 & x_{2,2} y_{1,1} - 2 x_{1,2} y_{1,2} + x_{1,1} y_{2,2} - y_{1,1} - y_{2,2} = 0.
\end{aligned}
\right.
\end{equation}

For the considered example, $\dim D_1 = 6$ and $\dim D_2 = 4$.
Note that for algebraic equations, the naïve approach consisting in calculating the dimension as the number of variables minus the number of equations (which is valid for systems of \emph{linear} equations) does not apply.
As we discuss in the next section, elaborate algorithms have to be used to compute these dimensions.

\section{Calculating the real dimension of an algebraic variety}

As it was discussed in the main text of the article, in order to find the number of parameters required for obtaining a BIC, we need to calculate the dimensions of the algebraic varieties $D_1$ and $D_2$.
The development of exact and efficient algorithms for calculating the real dimension of an algebraic variety remains an active field of research~\cite{Bannwarth:2015, Lairez:2021}.
A classic algorithm for solving algebraic equations (and finding the dimension of the solution set) is the cylindrical algebraic decomposition (CAD)~\cite{Basu:2006}.
Unfortunately, the computational complexity of the CAD is double exponential in the number of variables, which makes it inapplicable for the problems considered in the current paper.
Indeed, the number of variables and equations for the system of Eq.~\eqref{eqsupp:D2} amounts to~12 and~11, respectively.
For some examples considered in Table 1 of the main text of the paper, the number of variables reaches~100.

In this regard, we used the following \emph{numerical} approach for estimating the dimension.
To check whether the dimension is no less than~$M$, we fix the values of~$M$ randomly chosen variables to some random values and optimize the values of the remaining variables to satisfy the equations.
This approach was implemented with arbitrary-precision arithmetic with~100 decimal digits.
The solution was considered found when the discrepancies in the equations were less than~$10^{-75}$.
Such an approach enables efficiently estimating the dimensions of the algebraic equations appearing in the considered light scattering problem.
The time required for estimating the dimension varies from few seconds (for the equations of the previous section) to several hours (for the most complex structures considered in Tables~1 and~2 of the main text of the article).



\bibliographystyle{unsrt}
\bibliography{Dimension}

%
%
%

\end{document}